\documentclass{article}
\usepackage{slashed,verbatim,subfigure}
\usepackage[numbers,sort&compress]{natbib}
\usepackage{amsmath}
\usepackage{amssymb}
\usepackage{appendix}
\usepackage{graphicx}
\usepackage{dcolumn}
\usepackage{authblk}
\usepackage{bm}
\usepackage[colorlinks=true,linktocpage=true,
linkcolor=blue,citecolor=blue]{hyperref}
\usepackage[a4paper]{geometry}
\newcommand{\be}{\begin{eqnarray}}
\newcommand{\ee}{\end{eqnarray}}

\begin{document}
\large
\title{\bf{Studying the thermoelectric properties of an anisotropic QGP medium}}
\author{Shubhalaxmi Rath\thanks{shubhalaxmirath@gmail.com} }
\author{Nicol\'{a}s A. Neill\thanks{naneill@outlook.com}}
\affil{Centro Multidisciplinario de F\'isica, Vicerrector\'ia de Investigaci\'on, Universidad Mayor, 8580745 Santiago, Chile}
\date{}
\maketitle

\begin{abstract}
We have studied how the thermoelectric properties of the quark-gluon plasma (QGP) are affected by a weak-momentum anisotropy arising from the asymptotic expansion of matter in the initial stages of 
ultrarelativistic heavy-ion collisions. The highly energetic medium produced in such collisions exhibits 
a notable temperature difference between its central and peripheral regions. This temperature gradient induces an electric field whose magnitude per unit temperature gradient, in the limit of vanishing electric current, defines the Seebeck coefficient of the medium. We have calculated the Seebeck coefficients for both individual quark flavors and the entire QGP medium in the presence of expansion-induced anisotropy by solving the relativistic Boltzmann transport equation in the relaxation time approximation within the kinetic theory framework. The partonic interactions are incorporated through their effective thermal masses within the quasiparticle model for an anisotropic QGP medium. We have observed that the magnitude of the Seebeck coefficient for each quark flavor as well as for the entire QGP medium increases in the presence of expansion-induced anisotropy, indicating a stronger induced electric field in the anisotropic medium compared to the isotropic case. Given that an increase in the Seebeck coefficient may lead to observable signatures such as charge asymmetries in particle distributions and to modifications in the transport behavior of the QGP, these results may provide useful input for future phenomenological studies investigating the internal structure and phase properties of the QGP in heavy-ion collisions. 

\end{abstract}

\newpage

\section{Introduction}
A primary goal of the ultrarelativistic heavy ion collisions at the Relativistic Heavy Ion Collider (RHIC) 
and the Large Hadron Collider (LHC) is to study the properties of a new state of the strongly interacting matter, known as quark-gluon plasma (QGP). The early stages of ultrarelativistic heavy-ion collisions may possess a finite, though small, baryon asymmetry. While this asymmetry is typically modest, it can play a significant role in determining the thermodynamic behavior and the transport properties of the QGP medium. According to some studies, the baryon chemical potential may reach values as high as 300 MeV at temperatures around 160 MeV, indicating that the finite baryon density effects become relevant even at high energies \cite{P:JPG28'2002,Cleymans:JPG35'2008,Andronic:NPA837'2010}. Additionally, in the strong magnetic field regime, the baryon chemical potential has been reported to go up from 0.1 GeV to 0.6 GeV \cite{Fukushima:PRL117'2016}, implying an increase in the quark chemical potential. Various transport coefficients of the partonic medium have been found to be noticeably modulated by the cumulative effects of magnetic field and baryon asymmetry \cite{Rath:EPJC80'2020,Rath:EPJC81'2021,Rath:EPJC82'2022,Rath:EPJA59'2023}. Further, the matter produced in these collisions can develop a local momentum anisotropy due to the asymptotic free expansion of the fireball in the beam direction as compared to its transverse direction \cite{Dumitru:PLB662'2008,Dumitru:PRD79'2009}. This anisotropy, in combination with baryon asymmetry, can significantly alter the microscopic dynamics and macroscopic observables of the medium. Therefore, a comprehensive understanding of the QGP must consider the effects of both the expansion-induced anisotropy and finite baryon asymmetry. The momentum-space anisotropy in the QGP can be characterized by an anisotropy parameter $\xi$, which quantifies the deviation from isotropy in terms of the transverse ($p_T$) and longitudinal ($p_L$) momentum components. In this type of anisotropic configuration, the transverse component of momentum exceeds the longitudinal one, leading to a positive anisotropy parameter ($\xi=\langle {\bf p}_{T}^{2}\rangle/(2\langle p_{L}^{2}\rangle)-1$). When the anisotropy is weak ($\xi<1$), the parton distribution function can be approximated by a deformation of the isotropic distribution, effectively compressed along a preferred direction. Such anisotropic distributions modify the microscopic phase-space structure of the medium, and its transport properties. Within the kinetic theory framework, the transport coefficients are calculated using the parton distribution functions and their dispersion relations. The presence of anisotropy alters these inputs, thereby affecting the resulting transport coefficients. Recent studies have explored these effects in various contexts, emphasizing the need to explicitly incorporate anisotropy in the transport modeling to achieve a realistic description of the QGP in heavy-ion collisions \cite{Martinez:PRC78'2008,Thakur:PRD88'2013,Ryblewski:PRD92'2015,Mukherjee:EPJA53'2017,Rath:PRD100'2019,
Rath:PRD102'2020,Rath:EPJA62'2026}. 

In the present work, we explore the effect of the expansion-induced anisotropy on the thermoelectric properties by calculating the Seebeck coefficient for the QGP medium created in highly energetic heavy-ion collisions with a temperature gradient between the core and the peripheral regions of the fireball. We incorporate the effects of anisotropy, along with those of temperature and chemical potential in determining the effective thermal masses of partons within the quasiparticle model. These masses, which depend on the temperature, chemical potential, and anisotropy, are subsequently used to study the thermoelectric properties of the QGP medium. 

The QGP created in ultrarelativistic heavy-ion collisions features a hot central core surrounded by cooler peripheral regions. This spatial temperature gradient leads to a gradient in charge-carrier 
density, thereby inducing an electric field. This phenomenon of generation of an electric current due to a temperature gradient in a conducting medium is termed as the Seebeck effect, and its magnitude can be influenced by the bulk properties of the medium. When a temperature gradient exists within a conducting medium, the charge carriers with higher energies move from hotter region to cooler region until the induced electric field becomes strong enough to suppress further flow. The Seebeck coefficient measures the magnitude of the induced electric field per unit temperature gradient in a medium. This thermoelectric coefficient can be determined by imposing the condition of vanishing electric current \cite{Scheidemantel:PRB68'2003,Callen:BOOK'1985}. The emergence of a temperature gradient in the QGP medium can be treated as a small external perturbation to compute the thermoelectric coefficient by using the linear 
response theory. In the strongly interacting partonic medium, both positive and negative charge carriers participate in the charge transport process, unlike condensed matter systems, where the ions are stationary and only one kind of charge carrier participates in the transport process. For the observation of the thermoelectric effect, the strongly interacting matter must possess a temperature gradient as well as finite chemical potential. Due to the finite chemical potential, there is an imbalance between the numbers of particles and antiparticles, supporting the emergence of a net thermoelectric current in the medium. The sign of the Seebeck coefficient reveals the dominant charge carriers in the medium. A positive sign describes a thermoelectric current flowing from hotter zone to cooler zone and indicates the dominance of the positive charge carriers, whereas a negative sign indicates the dominance of the negative charge carriers. Since gluons are electrically neutral, they do not contribute to the thermoelectric effect in the QGP medium. 

While the Seebeck coefficient of the hot QCD medium under various medium modifications, including hadronic degrees of freedom, magnetic field-induced anisotropy and specific scattering mechanisms, has been 
investigated in several earlier studies \cite{Bhatt:PRD99'2019,Abhishek:EPJC82'2022,Shaikh:PRD111'2025,Dey:PRD102'2020,Kurian:PRD103'2021,
Zhang:EPJC81'2021,Dey:PRD104'2021,Khan:PRD107'2023,Singh:PRD110'2024}, the physical mechanisms and the theoretical frameworks considered therein differ substantially from the present work. In ref. \cite{Bhatt:PRD99'2019}, the Seebeck coefficient for the hadronic matter has been estimated using kinetic theory within the relaxation time approximation, where the calculation is performed for hadronic matter modeled by the hadron resonance gas model with hadrons and resonance states up to a cutoff in the mass as 2.25 GeV. Reference \cite{Abhishek:EPJC82'2022} focuses on the determination of the thermoelectric transport coefficients for quark matter within the ambit of the Nambu-Jona-Lasinio (NJL) model by using the relativistic Boltzmann transport equation within the relaxation time approximation, wherein the relaxation times for the quarks are estimated from the quark-quark and quark-antiquark scattering through meson exchange within the NJL model. Reference \cite{Shaikh:PRD111'2025} uses a novel relaxation time approximation model within the kinetic theory framework in investigating the Seebeck effect of the QGP medium. Reference \cite{Dey:PRD102'2020} has studied the Seebeck effect in a thermal QCD medium in the presence of a strong magnetic field, wherein the Seebeck coefficient has been determined using the relativistic Boltzmann transport equation in the relaxation time approximation in conjunction with the strong magnetic field limit. In references \cite{Kurian:PRD103'2021,Khan:PRD107'2023}, the thermoelectric behavior of the quark-gluon plasma has been studied in the presence of the different strengths of magnetic field, wherein the thermoelectric transport coefficients are calculated by utilizing the relaxation time approximation and Bhatnagar-Gross-Krook collision kernels in an effective Boltzmann equation. In particular, ref. \cite{Zhang:EPJC81'2021} analyzed the thermoelectric transport coefficients in a magnetized QGP medium under the weak and the strong magnetic field limits, where anisotropy primarily arises from the Landau quantization and magnetic field-induced phase-space modifications, and where the Nernst effect necessarily accompanies the Seebeck effect. Although ref. \cite{Zhang:EPJC81'2021} briefly mentions momentum-space anisotropy due to rapid longitudinal expansion, this anisotropy is incorporated only through deformed distribution functions with fixed masses within an already magnetized framework. In contrast, the present work focuses on a purely expansion-induced anisotropic QGP medium in the absence of any magnetic field, treating longitudinal expansion as the primary source of anisotropy, which is an intrinsic feature of relativistic heavy ion collisions. The anisotropy parameter enters self-consistently into the quasiparticle description, modifying the effective thermal masses and dispersion relations, and thereby directly influencing the thermoelectric transport properties. This allows us to isolate and quantify the impact of expansion-induced momentum anisotropy on the Seebeck coefficient, revealing a distinct and previously unexplored mechanism linking the early time QGP dynamics to the thermoelectric response. To the best of our knowledge, a self-consistent quasiparticle analysis of the Seebeck coefficient driven purely by expansion-induced anisotropy has not been reported previously. Reference \cite{Dey:PRD104'2021} has studied the thermoelectric response of a thermal QCD medium in the presence of a weak magnetic field, wherein the thermoelectric transport coefficients have been determined using the relativistic Boltzmann transport equation in the relaxation time approximation in association with the weak magnetic field limit. Reference \cite{Singh:PRD110'2024} has estimated the thermoelectric transport coefficients and the induced electric field in the presence of an external time-varying magnetic field, including the quantum effect of the Landau quantization, and explored the effects of the intensity and decay parameter of the magnetic field on the induced electric field. Moreover, unlike earlier works, the present analysis incorporates expansion-induced anisotropy self-consistently into the quasiparticle description, where the effective thermal masses of partons encode the interactions among them, and depend explicitly on the temperature, chemical potential and anisotropy parameter. This leads to modified dispersion relations, transport dynamics and distribution functions which are qualitatively different from those obtained in isotropic and magnetic field-induced anisotropic scenarios. As a result, the present work provides new insights into how the early time expansion dynamics of the quark-gluon plasma can significantly influence its thermoelectric response and the induced electric field, thereby extending the existing studies of the Seebeck effect in the strongly interacting matter. 

This paper is organized as follows. In section \ref{sec:thermo}, we describe the calculation of the Seebeck coefficient for an anisotropic QGP medium. Section \ref{sec:quasiparticle} discusses the quasiparticle 
model in the presence of expansion-induced anisotropy. The results of the aforementioned thermoelectric coefficient are discussed by considering the effective thermal masses of partons within the quasiparticle model in section \ref{sec:results}. Finally, section \ref{sec:conclusions} summarizes our conclusions. 

\section{Thermoelectric properties of an anisotropic QGP medium}\label{sec:thermo}
At early times in heavy-ion collisions, the produced matter may experience faster longitudinal 
expansion than transverse radial expansion, developing a local momentum anisotropy. For 
weak-momentum anisotropy ($\xi<1$) along a direction $\mathbf{n}$, the anisotropic distribution function closely resembles the isotropic form, with the tail of distribution being curtailed \cite{Romatschke:PRD68'2003}. Accordingly, the quark, antiquark, and gluon distribution functions are 
modified to 
\be\label{A.D.F.Q.}
&&f_f^\xi=\frac{N(\xi)}{e^{\beta\left(\sqrt{\omega_f^2+\xi(\mathbf{p}\cdot\mathbf{n})^2}-\mu_f\right)}+1} ~, \\ 
\label{A.D.F.A.}&&\bar{f}_f^\xi=\frac{N(\xi)}{e^{\beta\left(\sqrt{\omega_f^2+\xi(\mathbf{p}\cdot\mathbf{n})^2}+\mu_f\right)}+1} ~, \\ 
\label{A.D.F.G.}&&f_g^\xi=\frac{N(\xi)}{e^{\beta\sqrt{\omega_g^2+\xi(\mathbf{p}\cdot\mathbf{n})^2}}-1} 
~,\ee
where $N(\xi)=\sqrt{1+\xi}$ is a normalization factor \cite{Romatschke:PRD70'2004} and $\xi$ is the anisotropy parameter. Expanding to $\mathcal{O}(\xi)$, the distribution functions $f_f^\xi$, $\bar{f}_f^\xi$ and 
$f_g^\xi$ take the following forms, 
\be\label{E.Q.}
&&f_f^\xi=f_f+\frac{\xi f_f}{2}-\frac{\xi\beta(\mathbf{p}\cdot\mathbf{n})^2}{2\omega_f}f_f\left(1-f_f\right), \\ 
\label{E.A.}&&\bar{f}_f^\xi=\bar{f}_f+\frac{\xi \bar{f}_f}{2}-\frac{\xi\beta(\mathbf{p}\cdot\mathbf{n})^2}{2\omega_f}\bar{f}_f\left(1-\bar{f}_f\right), 
\\ 
\label{E.G.}&&f_g^\xi=f_g+\frac{\xi f_g}{2}-\frac{\xi\beta(\mathbf{p}\cdot\mathbf{n})^2}{2\omega_g}f_g\left(1+f_g\right)
.\ee
Here, $f_f$, $\bar{f}_f$ and $f_g$ denote the isotropic distribution functions, and are expressed as
\be\label{I.D.F.Q.}
&&f_f=\frac{1}{e^{\beta\left(\omega_f-\mu_f\right)}+1} ~, \\ 
&&\label{I.D.F.A.Q.}\bar{f}_f=\frac{1}{e^{\beta\left(\omega_f+\mu_f\right)}+1} ~, \\ 
&&\label{I.D.F.G.}f_g=\frac{1}{e^{\beta\omega_g}-1}
~,\ee
with $\beta=1/T$, $\mu_f$ the chemical potential, $\omega_f$ the energy of the $f$th flavor of quark (and antiquark), and $\omega_g$ the gluon energy in the QGP medium. The anisotropy can be quantified by the anisotropy parameter, which is generally defined in terms of the transverse and longitudinal momenta as
\be\label{parameter}
\xi=\frac{\left\langle\mathbf{p}_T^2\right\rangle}{2\left\langle p_L^2\right\rangle}-1
~,\ee
where $p_L=\mathbf{p}\cdot\mathbf{n}$, $\mathbf{p}_T=\mathbf{p}-\mathbf{n}\cdot(\mathbf{p}\cdot\mathbf{n})$, $\mathbf{p}\equiv(\rm{p}\sin\theta\cos\phi,\rm{p}\sin\theta\sin\phi,\rm{p}\cos\theta)$, $\mathbf{n}=(\sin\alpha,0,\cos\alpha)$, $\alpha$ is the angle between the $z$-axis and the anisotropy direction, $(\mathbf{p}\cdot\mathbf{n})^2=\rm{p}^2c(\alpha,\theta,\phi)=\rm{p}^2(\sin^2\alpha\sin^2\theta\cos^2\phi+\cos^2\alpha\cos^2\theta+\sin(2\alpha)\sin\theta\cos\theta\cos\phi)$. For $p_T \gg p_L$, the suppression of particles with large momentum components occurs along the $\mathbf{n}$ direction, reflecting the fact that the anisotropy parameter $\xi$ is positive in this regime. This behavior arises due to the faster longitudinal expansion as compared to the transverse expansion. The weak-momentum anisotropy due to the asymptotic expansion of matter can modulate the thermoelectric properties of the QGP medium, which can be quantified via the Seebeck coefficient. 

The presence of a temperature gradient makes the charge carriers to move from the higher temperature 
regions to the lower temperature regions, thus inducing an electric current. The spatial part of the 
induced four-current, {\em i.e.} the electric current density in an anisotropic QGP medium is 
written as
\begin{eqnarray}\label{current density}
\mathbf{J} = \sum_f g_f \int\frac{d^3\rm{p}}{(2\pi)^3\omega_f}\mathbf{p}\left[q_f\delta f^\xi_f+{\bar q_f}\delta \bar{f}^\xi_f\right]
,\end{eqnarray}
where $g_f$, $q_f$ ($\bar{q}_f$) and $\delta f^\xi_f$ ($\delta \bar{f}^\xi_f$) are the degeneracy factor, electric charge and infinitesimal change in the anisotropic quark (antiquark) distribution function 
of flavor $f$, respectively. The infinitesimal changes in the anisotropic quark and antiquark distribution functions due to the influence of electric field are defined as $\delta f_f^\xi=f_f^{\prime \xi}-f_f^\xi$ and $\delta \bar{f}_f^\xi=\bar{f}_f^{\prime \xi}-\bar{f}_f^\xi$, where $f_f^\xi$ and $\bar{f}_f^\xi$ denote the equilibrium distribution functions of the $f$th flavor quark and antiquark, respectively. The relativistic Boltzmann transport equation (RBTE) in the relaxation time approximation is expressed as
\be\label{R.B.T.E.}
p^\mu\frac{\partial f_f^{\prime\xi}}{\partial x^\mu}+q_f F^{\rho\sigma} 
p_\sigma \frac{\partial f_f^{\prime\xi}}{\partial p^\rho}=-\frac{p_\nu u^\nu}{\tau_f}\delta f_f^\xi
~.\ee
Here, $F^{\rho\sigma}$ represents the electromagnetic field strength tensor and $u^\nu$ denotes the four-velocity of fluid in the local rest frame. The relaxation time for quark (antiquark) of $f$th 
flavor, $\tau_f$ ($\tau_{\bar{f}}$) is given \cite{Hosoya:NPB250'1985} by
\begin{eqnarray}
\tau_{f(\bar{f})}=\frac{1}{5.1T\alpha_s^2\log\left(1/\alpha_s\right)\left[1+0.12(2N_f+1)\right]}
~.\end{eqnarray}
In order to observe the response to an electric field, we use the components of $F^{\rho\sigma}$ associated with only electric field. For an infinitesimal deviation of the phase-space distribution function from equilibrium, it is possible to use $f_f^{\prime\xi} \rightarrow f_f^\xi$ in the left-hand side of eq. \eqref{R.B.T.E.}. Additionally, for the steady-state condition, $\frac{\partial f_f^{\prime\xi}}{\partial t}=0$. Thus, the RBTE \eqref{R.B.T.E.} reduces to 
\be\label{R.B.T.E.(1)}
\mathbf{p}\cdot\frac{\partial f_f^\xi}{\partial \mathbf{r}}+q_f\mathbf{E}\cdot\mathbf{p}\frac{\partial f_f^\xi}{\partial p_0}+q_f p_0\mathbf{E}\cdot\frac{\partial f_f^\xi}{\partial \mathbf{p}}=-\frac{p_0}{\tau_f}\delta f_f^\xi
~.\ee
The partial derivatives appearing in the above equation are determined as
\be
\nonumber\frac{\partial f_f^\xi}{\partial \mathbf{r}}&=&\left[\beta^2(\omega_f-\mu_f)f_f\left(1-f_f\right)+\frac{\xi\beta^2}{2}(\omega_f-\mu_f)f_f\left(1-f_f\right)+\frac{\xi c(\alpha,\theta,\phi)\beta^2{\rm p}^2}{2\omega_f}f_f\left(1-f_f\right)\right. \\ && \left.\nonumber -\frac{\xi c(\alpha,\theta,\phi)\beta^3{\rm p}^2}{2\omega_f}(\omega_f-\mu_f)f_f\left(1-f_f\right)\right. \\ && \left.+\frac{\xi c(\alpha,\theta,\phi)\beta^3{\rm p}^2}{\omega_f}(\omega_f-\mu_f)f^2_f\left(1-f_f\right)\right]\bm{\nabla}T(\mathbf{r})
, \\ 
\nonumber\frac{\partial f_f^\xi}{\partial p_0} &=& -\beta f_f\left(1-f_f\right)-\frac{\xi\beta}{2}f_f\left(1-f_f\right)+\frac{\xi c(\alpha,\theta,\phi)\beta{\rm p}^2}{2\omega^2_f}f_f\left(1-f_f\right) \\ && +\frac{\xi c(\alpha,\theta,\phi)\beta^2{\rm p}^2}{2\omega_f}f_f\left(1-f_f\right)-\frac{\xi c(\alpha,\theta,\phi)\beta^2{\rm p}^2}{\omega_f}f^2_f\left(1-f_f\right)
, \\
\nonumber\frac{\partial f_f^\xi}{\partial \mathbf{p}} &=& -\frac{\beta\mathbf{p}}{\omega_f}f_f\left(1-f_f\right)-\frac{\xi\beta\mathbf{p}}{2\omega_f}f_f\left(1-f_f\right)-\frac{\xi c(\alpha,\theta,\phi)\beta\mathbf{p}}{\omega_f}f_f\left(1-f_f\right) \\ && \nonumber+\frac{\xi c(\alpha,\theta,\phi)\beta{\rm p}^2\mathbf{p}}{2\omega^3_f}f_f\left(1-f_f\right)+\frac{\xi c(\alpha,\theta,\phi)\beta^2{\rm p}^2\mathbf{p}}{2\omega^2_f}f_f\left(1-f_f\right) \\ && -\frac{\xi c(\alpha,\theta,\phi)\beta^2{\rm p}^2\mathbf{p}}{\omega^2_f}f^2_f\left(1-f_f\right)
,\ee
where $\bm{\nabla}T(\mathbf{r})$ represents the temperature gradient in the medium. Using the values of the previous partial derivatives in eq. \eqref{R.B.T.E.(1)} and solving, $\delta f_f^\xi$ for quark is obtained 
as
\be
\nonumber\delta f_f^\xi &=& -\frac{\mathbf{p}\tau_f\beta^2}{\omega_f}f_f\left(1-f_f\right)\left[\left(1+\frac{\xi}{2}\right)(\omega_f-\mu_f)+\frac{\xi c(\alpha,\theta,\phi)}{2}\left(\frac{{\rm p}^2}{\omega_f}+\frac{2\beta{\rm p}^2}{\omega_f}(\omega_f-\mu_f)f_f\right.\right. \\ && \left.\left.\nonumber -\frac{\beta{\rm p}^2}{\omega_f}(\omega_f-\mu_f)\right)\right]\bm{\nabla}T(\mathbf{r})+\frac{2\tau_f \beta q_f\mathbf{E}\cdot\mathbf{p}}{\omega_f}f_f\left(1-f_f\right)\left[1+\frac{\xi}{2}\right. \\ && \left.+\frac{\xi c(\alpha,\theta,\phi)}{2}\left(1+\frac{2\beta{\rm p}^2f_f}{\omega_f}-\frac{\beta\rm{p}^2}{\omega_f}-\frac{\rm{p}^2}{\omega_f^2}\right)\right]
.\ee
Similarly, $\delta \bar{f}_f^\xi$ for antiquark is determined as
\be
\nonumber\delta \bar{f}_f^\xi &=& -\frac{\mathbf{p}\tau_{\bar{f}}\beta^2}{\omega_f}\bar{f}_f\left(1-\bar{f}_f\right)\left[\left(1+\frac{\xi}{2}\right)(\omega_f+\mu_f)+\frac{\xi c(\alpha,\theta,\phi)}{2}\left(\frac{{\rm p}^2}{\omega_f}+\frac{2\beta{\rm p}^2}{\omega_f}(\omega_f+\mu_f)\bar{f}_f\right.\right. \\ && \left.\left.\nonumber -\frac{\beta{\rm p}^2}{\omega_f}(\omega_f+\mu_f)\right)\right]\bm{\nabla}T(\mathbf{r})+\frac{2\tau_{\bar{f}} \beta \bar{q}_f\mathbf{E}\cdot\mathbf{p}}{\omega_f}\bar{f}_f\left(1-\bar{f}_f\right)\left[1+\frac{\xi}{2}\right. \\ && \left.+\frac{\xi c(\alpha,\theta,\phi)}{2}\left(1+\frac{2\beta{\rm p}^2\bar{f}_f}{\omega_f}-\frac{\beta\rm{p}^2}{\omega_f}-\frac{\rm{p}^2}{\omega_f^2}\right)\right]
.\ee
Substituting the values of $\delta f_f^\xi$ and $\delta \bar{f}_f^\xi$ into eq. \eqref{current density} and performing the angular integration, the induced electric current density for a single quark flavor becomes
\begin{eqnarray}\label{current density (1)}
\nonumber\mathbf{J} &=& \frac{g_fq^2_f\beta}{3\pi^2}\int d{\rm p} ~ \frac{{\rm p}^4}{\omega_f^2}\left[\tau_f f_f\left(1-f_f\right)+\tau_{\bar{f}} \bar{f}_f\left(1-\bar{f}_f\right)\right]\left(1+\frac{\xi}{2}\right)\mathbf{E} \\ && \nonumber -\frac{\xi g_fq^2_f\beta}{9\pi^2}\int d{\rm p} ~ \frac{{\rm p}^4}{\omega_f^2}\left[\tau_f f_f\left(1-f_f\right)\left\lbrace \frac{{\rm p}^2}{2\omega_f^2}+\frac{{\rm p}^2\beta}{2\omega_f}-\frac{{\rm p}^2\beta f_f}{\omega_f}-\frac{1}{2} \right\rbrace+\tau_{\bar{f}} \bar{f}_f\left(1-\bar{f}_f\right)\right. \\ && \left.\nonumber\times\left\lbrace \frac{{\rm p}^2}{2\omega_f^2}+\frac{{\rm p}^2\beta}{2\omega_f}-\frac{{\rm p}^2\beta \bar{f}_f}{\omega_f}-\frac{1}{2} \right\rbrace\right]\mathbf{E}-\frac{g_fq_f\beta^2}{6\pi^2}\int d{\rm p} ~ \frac{{\rm p}^4}{\omega_f^2}\left[\tau_f(\omega_f-\mu_f)f_f\left(1-f_f\right)\right. \\ && \left.\nonumber -\tau_{\bar{f}}(\omega_f+\mu_f)\bar{f}_f\left(1-\bar{f}_f\right)\right]\left(1+\frac{\xi}{2}\right)\bm{\nabla}T(\mathbf{r})-\frac{\xi g_fq_f\beta^2}{18\pi^2}\int d{\rm p} ~ \frac{{\rm p}^4}{\omega_f^2}\left[\tau_f f_f\left(1-f_f\right)\left\lbrace \frac{{\rm p}^2}{2\omega_f}\right.\right. \\ && \left.\left.\nonumber -\frac{{\rm p}^2\beta}{2\omega_f}(\omega_f-\mu_f)+\frac{{\rm p}^2\beta f_f}{\omega_f}(\omega_f-\mu_f) \right\rbrace-\tau_{\bar{f}} \bar{f}_f\left(1-\bar{f}_f\right)\left\lbrace \frac{{\rm p}^2}{2\omega_f}-\frac{{\rm p}^2\beta}{2\omega_f}(\omega_f+\mu_f)\right.\right. \\ && \left.\left.+\frac{{\rm p}^2\beta \bar{f}_f}{\omega_f}(\omega_f+\mu_f) \right\rbrace\right]\bm{\nabla}T(\mathbf{r})
.\end{eqnarray}
To determine the electric field induced by the temperature gradient, we assume that the electric current density vanishes ($\mathbf{J}=0$). This relates the temperature gradient to the induced electric field 
as
\begin{eqnarray}\label{T.E.}
\mathbf{E}=S \bm{\nabla}T(\mathbf{r})
,\end{eqnarray}
where $S$ is the Seebeck coefficient, defined as the induced electric field per unit temperature gradient at zero electric current. Equating eq. \eqref{current density (1)} to zero, we get 
\begin{eqnarray}\label{T.E.(1)}
\mathbf{E}=\left[\frac{\frac{\beta}{2}L^2_f+\frac{\xi\beta}{4}M^{21}_f+\frac{\xi\beta}{12}M^{22}_f-\frac{\xi\beta^2}{12}M^{23}_f+\frac{\xi\beta^2}{6}M^{24}_f}{q_fL^1_f+\frac{\xi q_f}{2}M^{11}_f-\frac{\xi q_f}{6}M^{12}_f-\frac{\xi\beta q_f}{6}M^{13}_f+\frac{\xi\beta q_f}{3}M^{14}_f+\frac{\xi q_f}{6}M^{15}_f}\right]\bm{\nabla}T(\mathbf{r})
,\end{eqnarray}
with the following integral coefficients, 
\begin{eqnarray}
&&L^1_f=\int d{\rm p} ~ \frac{{\rm p}^4}{\omega_f^2}\left[\tau_f f_f\left(1-f_f\right)+\tau_{\bar{f}} \bar{f}_f\left(1-\bar{f}_f\right)\right], \\ 
&&M^{11}_f=\int d{\rm p} ~ \frac{{\rm p}^4}{\omega_f^2}\left[\tau_f f_f\left(1-f_f\right)+\tau_{\bar{f}} \bar{f}_f\left(1-\bar{f}_f\right)\right], \\ 
&&M^{12}_f=\int d{\rm p} ~ \frac{{\rm p}^6}{\omega_f^4}\left[\tau_f f_f\left(1-f_f\right)+\tau_{\bar{f}} \bar{f}_f\left(1-\bar{f}_f\right)\right], \\ 
&&M^{13}_f=\int d{\rm p} ~ \frac{{\rm p}^6}{\omega_f^3}\left[\tau_f f_f\left(1-f_f\right)+\tau_{\bar{f}} \bar{f}_f\left(1-\bar{f}_f\right)\right], \\ 
&&M^{14}_f=\int d{\rm p} ~ \frac{{\rm p}^6}{\omega_f^3}\left[\tau_f f^2_f\left(1-f_f\right)+\tau_{\bar{f}} \bar{f}^2_f\left(1-\bar{f}_f\right)\right], \\ 
&&M^{15}_f=\int d{\rm p} ~ \frac{{\rm p}^4}{\omega_f^2}\left[\tau_f f_f\left(1-f_f\right)+\tau_{\bar{f}} \bar{f}_f\left(1-\bar{f}_f\right)\right], \\ 
&&L^{2}_f=\int d{\rm p} ~ \frac{{\rm p}^4}{\omega_f^2}\left[\tau_f(\omega_f-\mu_f)f_f\left(1-f_f\right)-\tau_{\bar{f}}(\omega_f+\mu_f)\bar{f}_f\left(1-\bar{f}_f\right)\right], \\ 
&&M^{21}_f=\int d{\rm p} ~ \frac{{\rm p}^4}{\omega_f^2}\left[\tau_f(\omega_f-\mu_f)f_f\left(1-f_f\right)-\tau_{\bar{f}}(\omega_f+\mu_f)\bar{f}_f\left(1-\bar{f}_f\right)\right], \\ 
&&M^{22}_f=\int d{\rm p} ~ \frac{{\rm p}^6}{\omega_f^3}\left[\tau_ff_f\left(1-f_f\right)-\tau_{\bar{f}}\bar{f}_f\left(1-\bar{f}_f\right)\right], \\ 
&&M^{23}_f=\int d{\rm p} ~ \frac{{\rm p}^6}{\omega_f^3}\left[\tau_f(\omega_f-\mu_f)f_f\left(1-f_f\right)-\tau_{\bar{f}}(\omega_f+\mu_f)\bar{f}_f\left(1-\bar{f}_f\right)\right], \\ 
&&M^{24}_f=\int d{\rm p} ~ \frac{{\rm p}^6}{\omega_f^3}\left[\tau_f(\omega_f-\mu_f)f^2_f\left(1-f_f\right)-\tau_{\bar{f}}(\omega_f+\mu_f)\bar{f}^2_f\left(1-\bar{f}_f\right)\right]
.\end{eqnarray}
Comparing eq. \eqref{T.E.(1)} with eq. \eqref{T.E.}, the Seebeck coefficient for a single quark flavor in an anisotropic QGP medium is obtained as
\begin{eqnarray}\label{S.C.}
S_f=\frac{\frac{\beta}{2}L^2_f+\frac{\xi\beta}{4}M^{21}_f+\frac{\xi\beta}{12}M^{22}_f-\frac{\xi\beta^2}{12}M^{23}_f+\frac{\xi\beta^2}{6}M^{24}_f}{q_fL^1_f+\frac{\xi q_f}{2}M^{11}_f-\frac{\xi q_f}{6}M^{12}_f-\frac{\xi\beta q_f}{6}M^{13}_f+\frac{\xi\beta q_f}{3}M^{14}_f+\frac{\xi q_f}{6}M^{15}_f}
.\end{eqnarray}

After calculating the Seebeck coefficient for a thermal medium composed of a single particle species, we now consider a more realistic multicomponent system with $u$, $d$, and $s$ quark flavors. It is 
important to note that, since gluons are electrically neutral, they do not contribute to the thermoelectric current. The total electric current is the vector sum of the electric currents generated by each quark flavor and is written as
\begin{eqnarray}\label{Total current density}
\nonumber\mathbf{J} &=& \sum_f J_f \\ &=& \nonumber\sum_f\frac{g_fq^2_f\beta}{3\pi^2}\int d{\rm p} ~ \frac{{\rm p}^4}{\omega_f^2}\left[\tau_f f_f\left(1-f_f\right)+\tau_{\bar{f}} \bar{f}_f\left(1-\bar{f}_f\right)\right]\left(1+\frac{\xi}{2}\right)\mathbf{E} \\ && \nonumber -\sum_f\frac{\xi g_fq^2_f\beta}{9\pi^2}\int d{\rm p} ~ \frac{{\rm p}^4}{\omega_f^2}\left[\tau_f f_f\left(1-f_f\right)\left\lbrace \frac{{\rm p}^2}{2\omega_f^2}+\frac{{\rm p}^2\beta}{2\omega_f}-\frac{{\rm p}^2\beta f_f}{\omega_f}-\frac{1}{2} \right\rbrace+\tau_{\bar{f}} \bar{f}_f\left(1-\bar{f}_f\right)\right. \\ && \left.\nonumber\times\left\lbrace \frac{{\rm p}^2}{2\omega_f^2}+\frac{{\rm p}^2\beta}{2\omega_f}-\frac{{\rm p}^2\beta \bar{f}_f}{\omega_f}-\frac{1}{2} \right\rbrace\right]\mathbf{E}-\sum_f\frac{g_fq_f\beta^2}{6\pi^2}\int d{\rm p} ~ \frac{{\rm p}^4}{\omega_f^2}\left[\tau_f(\omega_f-\mu_f)f_f\left(1-f_f\right)\right. \\ && \left.\nonumber -\tau_{\bar{f}}(\omega_f+\mu_f)\bar{f}_f\left(1-\bar{f}_f\right)\right]\left(1+\frac{\xi}{2}\right)\bm{\nabla}T(\mathbf{r})-\sum_f\frac{\xi g_fq_f\beta^2}{18\pi^2}\int d{\rm p} ~ \frac{{\rm p}^4}{\omega_f^2}\left[\tau_f f_f\left(1-f_f\right)\left\lbrace \frac{{\rm p}^2}{2\omega_f}\right.\right. \\ && \left.\left.\nonumber -\frac{{\rm p}^2\beta}{2\omega_f}(\omega_f-\mu_f)+\frac{{\rm p}^2\beta f_f}{\omega_f}(\omega_f-\mu_f) \right\rbrace-\tau_{\bar{f}} \bar{f}_f\left(1-\bar{f}_f\right)\left\lbrace \frac{{\rm p}^2}{2\omega_f}-\frac{{\rm p}^2\beta}{2\omega_f}(\omega_f+\mu_f)\right.\right. \\ && \left.\left.+\frac{{\rm p}^2\beta \bar{f}_f}{\omega_f}(\omega_f+\mu_f) \right\rbrace\right]\bm{\nabla}T(\mathbf{r})
.\end{eqnarray}
By setting $\mathbf{J}=0$ in eq. \eqref{Total current density}, the total induced electric field reads 
\begin{eqnarray}\label{T.E.(2)}
\nonumber\mathbf{E} &=& \left[\left\lbrace\frac{\beta}{2}\sum_f q_fL^2_f+\frac{\xi\beta}{4}\sum_f q_fM^{21}_f+\frac{\xi\beta}{12}\sum_f q_fM^{22}_f-\frac{\xi\beta^2}{12}\sum_f q_fM^{23}_f\right.\right. \\ && \left.\left.\nonumber+\frac{\xi\beta^2}{6}\sum_f q_fM^{24}_f\right\rbrace\Big/\left\lbrace\sum_f q^2_fL^1_f+\frac{\xi}{2}\sum_f q^2_fM^{11}_f-\frac{\xi}{6}\sum_f q^2_fM^{12}_f\right.\right. \\ && \left.\left.-\frac{\xi\beta}{6}\sum_f q^2_fM^{13}_f+\frac{\xi\beta}{3}\sum_f q^2_fM^{14}_f+\frac{\xi}{6}\sum_f q^2_fM^{15}_f\right\rbrace\right]\bm{\nabla}T(\mathbf{r})
,\end{eqnarray}
where we have used the fact that, all quark flavors share the same degeneracy factor ($g_f$). Comparing equations \eqref{T.E.(2)} and \eqref{T.E.}, the total Seebeck coefficient for an anisotropic QGP medium 
is obtained as
\begin{eqnarray}\label{S.C.(1)}
\nonumber S &=& \left[\frac{\beta}{2}\sum_f q_fL^2_f+\frac{\xi\beta}{4}\sum_f q_fM^{21}_f+\frac{\xi\beta}{12}\sum_f q_fM^{22}_f-\frac{\xi\beta^2}{12}\sum_f q_fM^{23}_f\right. \\ && \left.\nonumber+\frac{\xi\beta^2}{6}\sum_f q_fM^{24}_f\right]\Big/\left[\sum_f q^2_fL^1_f+\frac{\xi}{2}\sum_f q^2_fM^{11}_f-\frac{\xi}{6}\sum_f q^2_fM^{12}_f\right. \\ && \left.-\frac{\xi\beta}{6}\sum_f q^2_fM^{13}_f+\frac{\xi\beta}{3}\sum_f q^2_fM^{14}_f+\frac{\xi}{6}\sum_f q^2_fM^{15}_f\right]
.\end{eqnarray}
Equivalently, the total Seebeck coefficient of the anisotropic QGP medium can be expressed in terms of the single-flavor Seebeck coefficients as
\begin{eqnarray}
S=\frac{\sum_f q^2_f S_f\left[L^1_f+\frac{\xi}{2}M^{11}_f-\frac{\xi}{6}M^{12}_f-\frac{\xi\beta}{6}M^{13}_f+\frac{\xi\beta}{3}M^{14}_f+\frac{\xi}{6}M^{15}_f\right]}{\sum_f q^2_f\left[L^1_f+\frac{\xi}{2}M^{11}_f-\frac{\xi}{6}M^{12}_f-\frac{\xi\beta}{6}M^{13}_f+\frac{\xi\beta}{3}M^{14}_f+\frac{\xi}{6}M^{15}_f\right]}
.\end{eqnarray}

\section{Quasiparticle description of QGP medium in the presence of expansion-induced anisotropy}\label{sec:quasiparticle}
In the quasiparticle model, each constituent of QGP medium acquires a thermally generated mass due to its interaction with the surrounding medium, thus elucidating the collective properties of the medium. The quasiparticle model takes into account both the thermal and current masses of partons, treating the QGP as a medium of noninteracting quasipartons with thermally generated masses. This approach was originally introduced by Goloviznin and Satz \cite{Goloviznin:ZPC57'1993} to study the gluonic plasma, and later extended by Peshier {\em et al.} \cite{Peshier:PRD54'1996,Peshier:PRD66'2002} to investigate the QGP equation of state using lattice QCD results at finite temperature. Further studies \cite{Bluhm:PLB620'2005,Bluhm:PRC76'2007} 
incorporated the temperature- and chemical-potential-dependent quasiparticle masses to explicate the lattice results. These studies collectively demonstrate that the high temperature QGP phase can be effectively described by using a thermodynamically consistent quasiparticle model. 

In the present work, we have used the quasiparticle model, where the effective mass of the $f$th quark flavor is given \cite{Bannur:EPJC50'2007,Bannur:JHEP0709'2007} by
\begin{equation}\label{E.M.}
m_{f}^2=m^2_{fC}+\sqrt{2}m_{fC} m_{fT}+m^2_{fT}
,\end{equation}
where $m_{fC}$ and $m_{fT}$ represent the current mass and the thermal mass of the quark, 
respectively. The thermal mass of a parton can get significantly modified by the 
presence of different extreme conditions and anisotropies. In the semiclassical transport theory, the thermal mass (squared) of a quark in a baryon asymmetric hot QCD medium is defined \cite{Kelly:PRD50'1994,Litim:PR364'2002} as
\be\label{Q.P.M.Q.(definition of quark mass)}
\nonumber m_{fT}^2 &=& \frac{g^2\left(N_c^2-1\right)}{2N_c}\int\frac{d^3{\rm p}}{(2\pi)^3} ~ \frac{1}{\rm p}\left[f_g+\frac{1}{2}\left(f_f+\bar{f}_f\right)\right] \\ &=& \frac{2g^2}{3\pi^2}\int d{\rm p} ~ {\rm p}\left[f_g+\frac{1}{2}\left(f_f+\bar{f}_f\right)\right]
.\ee
This equation \eqref{Q.P.M.Q.(definition of quark mass)} can be evaluated within the hard thermal loop (HTL) approximation. The simplified form of the thermal mass (squared) of quark up to one-loop is given \cite{Braaten:PRD45'1992,Peshier:PRD66'2002,Blaizot:PRD72'2005,Berrehrah:PRC89'2014} by
\be\label{Q.P.M.(Quark mass)}
m_{fT}^2=\frac{g^2T^2}{6}\left(1+\frac{\mu_f^2}{\pi^2T^2}\right)
,\ee
which depends on both temperature and chemical potential. Equation \eqref{E.M.} contains the above expression of $m_{fT}^2$ and the corresponding dispersion relation for quark within the quasiparticle description of an isotropic QGP medium with finite baryon asymmetry is written as
\be
&&\label{I.D.R.(Quark)}\omega_f=\sqrt{{\rm p}^2+m_{f}^2}
~.\ee
In this regime, the quark distribution function involves the abovementioned dispersion relation. In this work, we assume equal chemical potentials for $u$, $d$ and $s$ quark flavors, {\em i.e.} $\mu_f=\mu$. In eq. \eqref{Q.P.M.(Quark mass)}, $g$ denotes the one-loop running coupling, which is given \cite{Kapusta:BOOK'2006} by
\begin{eqnarray}
g^2=4\pi\alpha_s=\frac{48\pi^2}{\left(11N_c-2N_f\right)\ln\left({\Lambda^2}/{\Lambda_{\rm\overline{MS}}^2}\right)}
~,\end{eqnarray}
where $\Lambda_{\rm\overline{MS}}=0.176$ GeV, $\Lambda=2\pi\sqrt{T^2+\mu_f^2/\pi^2}$ for electrically charged particles (quarks and antiquarks) and $\Lambda=2 \pi T$ for gluons. 

The presence of expansion-induced anisotropy can significantly modify the thermal masses of partons and, consequently, their dispersion relations. Thus, in the presence of expansion-induced anisotropy, the 
effective mass of the $f$th flavor quark becomes
\begin{equation}\label{E.M.(Anisotropy)}
m_{f\xi}^2=m^2_{fC}+\sqrt{2}m_{fC} m_{fT\xi}+m^2_{fT\xi}
,\end{equation}
where $m_{fT\xi}$ denotes the thermal mass of the quark in the anisotropic medium. In this regime, eq. \eqref{Q.P.M.Q.(definition of quark mass)} gets modified as
\be\label{Q.P.M.Q.(Anisotropy)}
m_{fT\xi}^2 &=& \frac{g^2\left(N_c^2-1\right)}{2N_c}\int\frac{d^3{\rm p}}{(2\pi)^3} ~ \frac{1}{\rm p}\left[f_g^\xi+\frac{1}{2}\left(f_f^\xi+\bar{f}_f^\xi\right)\right]
.\ee
It should be noted that the above equation defines the thermal mass (squared) of a quark in the presence of expansion-induced anisotropy within the kinetic theory framework and is not intended to represent the HTL screening mass derived from the full anisotropic self-energy. Substituting the values of the anisotropic distribution functions $f_f^\xi$, $\bar{f}_f^\xi$ and $f_g^\xi$ in eq. \eqref{Q.P.M.Q.(Anisotropy)} and calculating, we obtain the squared thermal mass of a quark in an expansion-induced anisotropic QGP medium as
\be\label{Q.P.M.Q.(Anisotropy 1)}
\nonumber m_{fT\xi}^2 &=& \frac{2g^2}{3\pi^2}\int d{\rm p} ~ {\rm p} \left[f_g+\frac{1}{2}\left(f_f+\bar{f}_f\right)\right]+\frac{\xi g^2}{3\pi^2}\int d{\rm p} ~ {\rm p} \left[f_g+\frac{1}{2}\left(f_f+\bar{f}_f\right)\right] \\ && -\frac{\xi\beta g^2}{9\pi^2}\int d{\rm p} ~ {\rm p^2} ~ f_g\left(1+f_g\right)-\frac{\xi\beta g^2}{18\pi^2}\int d{\rm p} ~ \frac{\rm p^3}{\omega_f}\left[f_f\left(1-f_f\right)+\bar{f}_f\left(1-\bar{f}_f\right)\right]
,\ee
where the first term on the right-hand side corresponds to the thermal mass (squared) of quark ($m_{fT}^2$) in an isotropic QGP medium. Hence, we have 
\be\label{Q.P.M.Q.(Anisotropic medium)}
\nonumber m_{fT\xi}^2 &=& m_{fT}^2+\frac{\xi g^2}{3\pi^2}\int d{\rm p} ~ {\rm p} \left[f_g+\frac{1}{2}\left(f_f+\bar{f}_f\right)\right]-\frac{\xi\beta g^2}{9\pi^2}\int d{\rm p} ~ {\rm p^2} ~ f_g\left(1+f_g\right) \\ && -\frac{\xi\beta g^2}{18\pi^2}\int d{\rm p} ~ \frac{\rm p^3}{\omega_f}\left[f_f\left(1-f_f\right)+\bar{f}_f\left(1-\bar{f}_f\right)\right]
.\ee
Equation \eqref{Q.P.M.Q.(Anisotropic medium)} reveals that, for an expansion-induced anisotropic medium with finite baryon asymmetry, the thermal mass of quark is modulated by the anisotropy parameter, in addition to 
its dependence on the temperature and chemical potential. Equation \eqref{E.M.(Anisotropy)} includes the above expression of $m_{fT\xi}^2$. Accordingly, the dispersion relation for quark gets modified and takes the following form within the quasiparticle description of an anisotropic QGP medium with finite baryon asymmetry, 
\be
&&\label{A.D.R.(Quark)}\omega_f=\sqrt{{\rm p}^2+m_{f\xi}^2}
~.\ee
In this regime, the quark distribution function incorporates the abovementioned dispersion relation. 

\begin{figure}[]
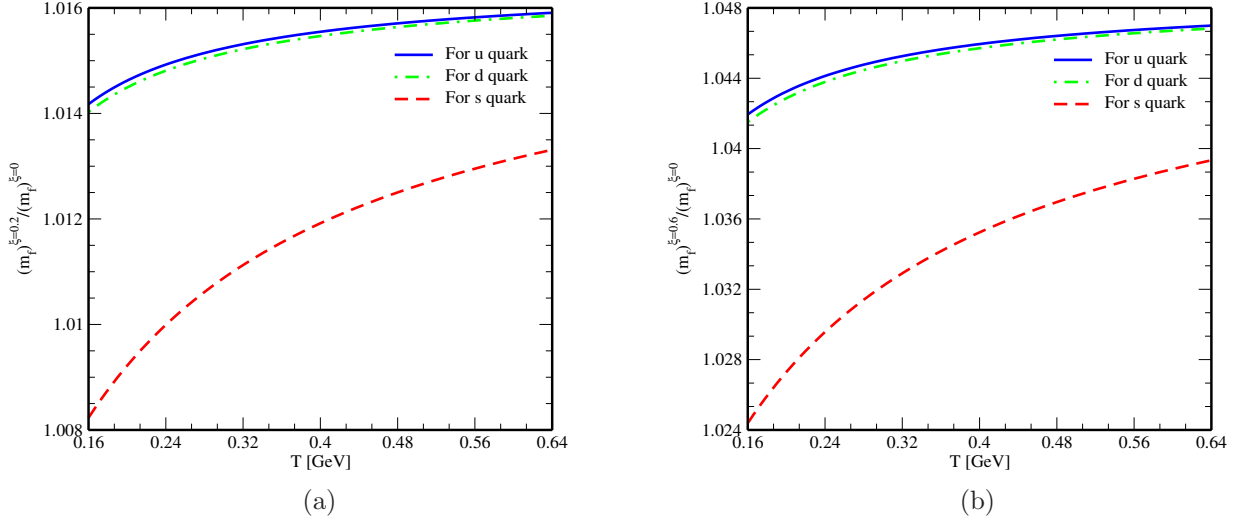

\begin{center}
\begin{tabular}{c c}
\includegraphics[width=7.4cm]{mq2.eps}&
\hspace{0.73 cm}
\includegraphics[width=7.4cm]{mq6.eps} \\
\ \ \ \ \ \ \ (a) & \ \ \ \  \ \ \ \ \ \ \ \ \ \ (b)
\end{tabular}
\caption{Variation of the quasiparticle mass of quark, normalized to its value in an isotropic medium, as a function of temperature for (a) $\xi=0.2$ and (b) $\xi=0.6$.}\label{Fig.m}
\end{center}
\end{figure}

\begin{figure}[]
\begin{center}
\begin{tabular}{c c}
\includegraphics[width=7.4cm]{dftu.eps}&
\hspace{0.73 cm}
\includegraphics[width=7.4cm]{dftd.eps} \\
\ \ \ \ \ \ \ (a) & \ \ \ \  \ \ \ \ \ \ \ \ \ \ (b)
\end{tabular}
\caption{Temperature dependence of the (a) $u$ quark and (b) $d$ quark distribution functions, normalized to their isotropic counterparts, for different values of the anisotropy parameter.}\label{Fig.df1}
\end{center}
\end{figure}

\begin{figure}[]
\begin{center}
\begin{tabular}{c c}
\includegraphics[width=7.4cm]{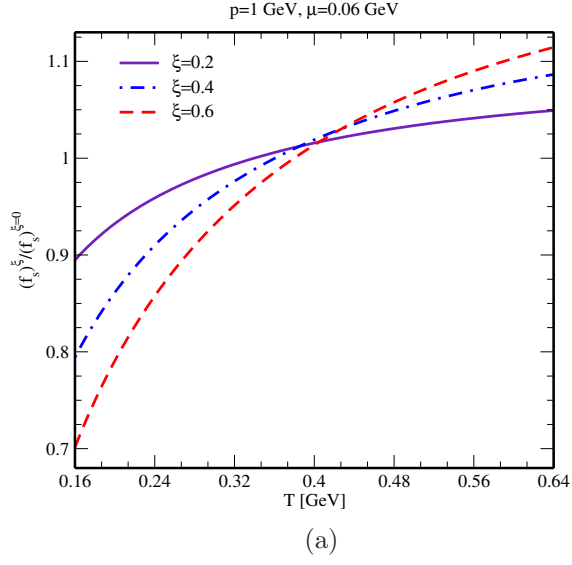}&
\hspace{0.73 cm}
\includegraphics[width=7.4cm]{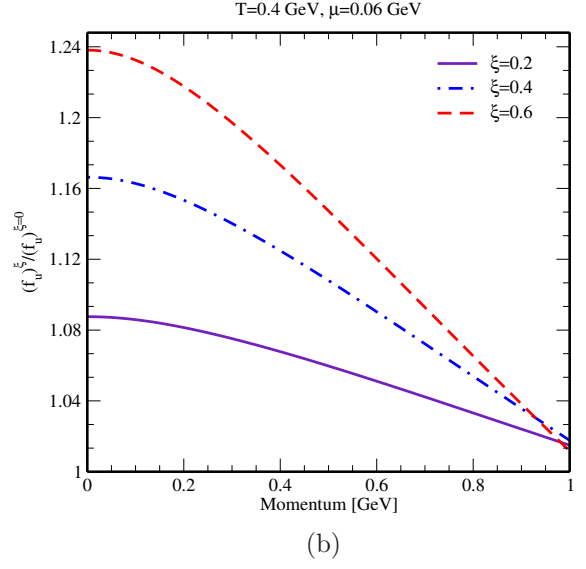} \\
\ \ \ \ \ \ \ (a) & \ \ \ \  \ \ \ \ \ \ \ \ \ \ (b)
\end{tabular}
\caption{(a) Temperature dependence of the $s$ quark distribution function and (b) momentum dependence of the $u$ quark distribution function, both normalized to their isotropic counterparts, for different values of the anisotropy parameter.}\label{Fig.df2}
\end{center}
\end{figure}

\begin{figure}[]
\begin{center}
\begin{tabular}{c c}
\includegraphics[width=7.4cm]{dfpd.eps}&
\hspace{0.73 cm}
\includegraphics[width=7.4cm]{dfps.eps} \\
\ \ \ \ \ \ \ (a) & \ \ \ \  \ \ \ \ \ \ \ \ \ \ (b)
\end{tabular}
\caption{Momentum dependence of the (a) $d$ quark and (b) $s$ quark distribution functions, normalized to their isotropic counterparts, for different values of the anisotropy parameter.}\label{Fig.df3}
\end{center}
\end{figure}

\begin{figure}[]
\begin{center}
\begin{tabular}{c c}
\includegraphics[width=7.4cm]{nd2.eps}&
\hspace{0.73 cm}
\includegraphics[width=7.4cm]{nd6.eps} \\
\ \ \ \ \ \ \ (a) & \ \ \ \  \ \ \ \ \ \ \ \ \ \ (b)
\end{tabular}
\caption{Temperature dependence of the number densities of different quark flavors, normalized to their isotropic counterparts for (a) $\xi=0.2$ and (b) $\xi=0.6$.}\label{Fig.nd}
\end{center}
\end{figure}

Figure \ref{Fig.m} shows the temperature dependence of the quasiparticle masses of $u$, $d$, and $s$ 
quarks, normalized to their corresponding isotropic counterparts, for different values of the 
anisotropy parameter. A moderate enhancement in the quasiparticle masses of quarks is observed with increasing anisotropy (figures \ref{Fig.m}(a) and \ref{Fig.m}(b)). Thus, the quarks become slightly more massive in an expansion-induced anisotropic medium, leading to modified dispersion relations. It is worth noting that, this anisotropy-induced enhancement of the quasiparticle effective mass is not in contradiction with earlier observations of reduced screening effect in an anisotropic plasma, since the latter is governed by the long-wavelength behavior of the gluon propagator (Debye screening mass), whereas the effective mass encodes medium-induced self-energy corrections entering the partonic dispersion relations. As the temperature increases, the quasiparticle masses for $u$, $d$ and $s$ quarks depart from their isotropic values. Further, the parton distribution functions get altered due to the impact of anisotropy via the dispersion relations. The resulting changes in the $u$-, $d$-, and $s$-quark distribution functions are displayed in figures \ref{Fig.df1}, \ref{Fig.df2} and \ref{Fig.df3}. In particular, the temperature dependence of the distribution functions is shown in figures \ref{Fig.df1}(a), \ref{Fig.df1}(b) and \ref{Fig.df2}(a) for different values of anisotropy parameter, while their momentum dependence is depicted in figures \ref{Fig.df2}(b), \ref{Fig.df3}(a) and \ref{Fig.df3}(b). It is observed that, for all quark flavors, anisotropy suppresses the distribution functions up to a certain temperature $T \approx 0.4$ GeV and enhances them at higher temperatures (figures \ref{Fig.df1}(a), \ref{Fig.df1}(b) and \ref{Fig.df2}(a)). Thus the deviation of distribution functions for different quark flavors from their isotropic counterparts increases beyond this temperature. However, no such turning point in momentum (figures \ref{Fig.df2}(b), \ref{Fig.df3}(a) and \ref{Fig.df3}(b)) is found, instead, anisotropy enhances the parton distribution functions across the entire momentum range. Consequently, the deviation of the parton number densities from their isotropic values becomes more significant with the increase of anisotropy. It is seen from figure \ref{Fig.nd} that, this deviation persists throughout the considered temperature range. We note that, although the normalization factor $N(\xi)$ preserves the number density for the exact anisotropic distribution, the weak-anisotropy expansion employed in the present work does not maintain this constraint order by order in $\xi$, leading to finite corrections to the number density at linear order. Additionally, the deviations of the quark distribution functions from their isotropic counterparts decrease with increasing momentum. 

Overall, the influence of anisotropy on the parton distribution functions is more pronounced at 
high temperatures and low momenta. In the kinetic theory framework, the parton distribution 
functions play a crucial role in determining the transport properties of the QGP medium, 
including the thermoelectric coefficient. Consequently, understanding their behavior under anisotropy provides valuable insight into how expansion-induced anisotropy influences the thermoelectric response of the QGP medium. 

\section{Results and discussions}\label{sec:results}
\begin{figure}[]
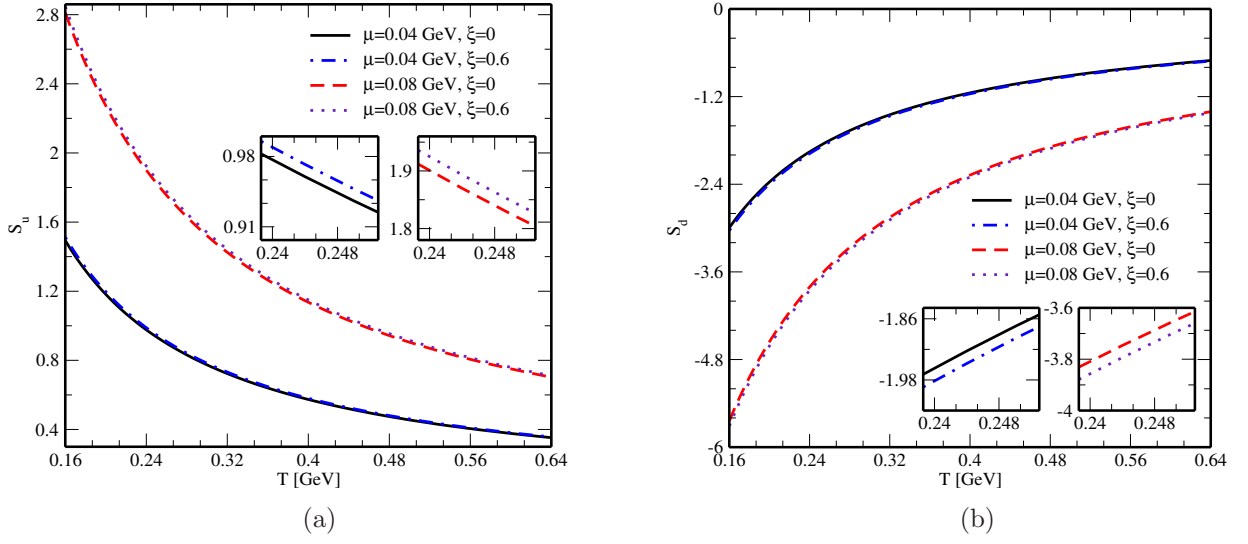

\begin{center}
\begin{tabular}{c c}
\includegraphics[width=7.4cm]{suaniso.eps}&
\hspace{0.73 cm}
\includegraphics[width=7.4cm]{sdaniso.eps} \\
\ \ \ \ \ \ \ (a) & \ \ \ \  \ \ \ \ \ \ \ \ \ \ (b)
\end{tabular}
\caption{Variation of Seebeck coefficient as a function of temperature for (a) $u$ quark and (b) $d$ quark.}\label{Fig.1}
\end{center}
\end{figure}

\begin{figure}[]
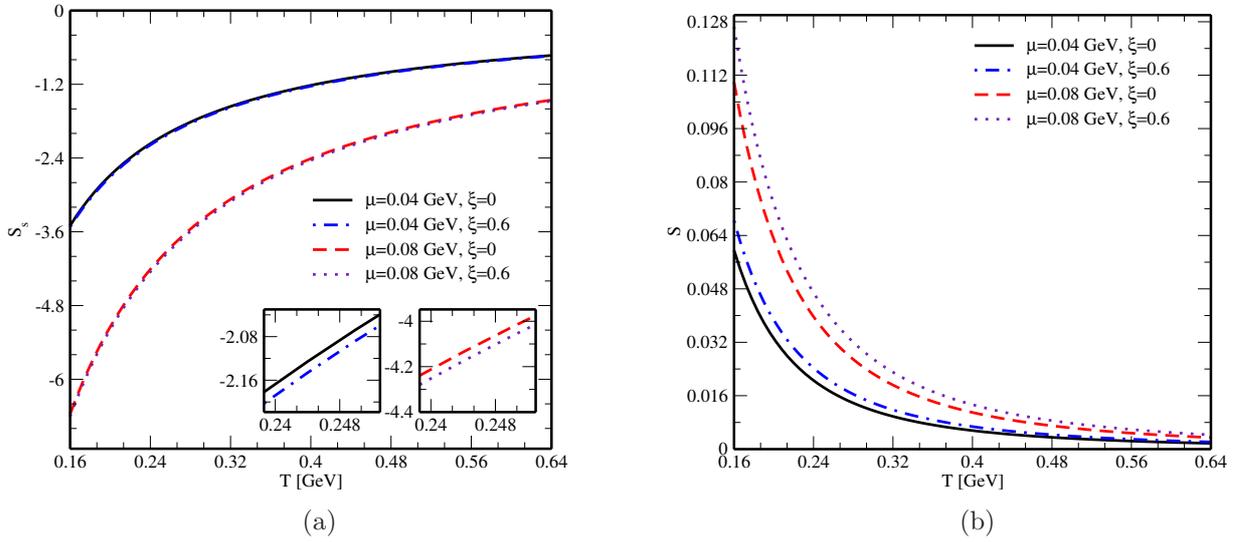

\begin{center}
\begin{tabular}{c c}
\includegraphics[width=7.4cm]{ssaniso.eps}&
\hspace{0.73 cm}
\includegraphics[width=7.4cm]{saniso.eps} \\
\ \ \ \ \ \ \ (a) & \ \ \ \  \ \ \ \ \ \ \ \ \ \ (b)
\end{tabular}
\caption{(a) Variation of Seebeck coefficient as a function of temperature for $s$ quark and (b) variation of the total Seebeck coefficient as a function of temperature.}\label{Fig.2}
\end{center}
\end{figure}

Figures \ref{Fig.1}(a), \ref{Fig.1}(b) and \ref{Fig.2}(a) show the effect of expansion-induced anisotropy on the Seebeck coefficients of individual quark flavors, while figure \ref{Fig.2}(b) displays the corresponding 
behavior of the total Seebeck coefficient ($S$) for the QGP medium. It is found that the magnitudes of 
Seebeck coefficients of individual quark flavors as well as the composite medium decrease as the temperature increases, while keeping the chemical potential and the anisotropy parameter fixed. On the other hand, the magnitudes of the Seebeck coefficients amplify with the increase of chemical potential at fixed values of the temperature and the anisotropy parameter. This is understood from the fact that the increasing chemical potential enhances the number of particles relative to antiparticles. Our observation shows an enhancement in the magnitudes of Seebeck coefficients of individual quark flavors as well as the whole QGP medium when the medium exhibits expansion-induced anisotropy. The influence of anisotropy on the Seebeck coefficients of individual quark flavors is moderate. However, the impact of anisotropy on the total Seebeck coefficient is found to be more evident than that of the individual quark flavors. The Seebeck coefficient for $u$ quark ($S_u$) remains positive throughout the considered temperature range, which is mainly due to its positive electric charge. On the other hand, the Seebeck coefficients for $d$ ($S_d$) and $s$ ($S_s$) quarks exhibit a sign reversal due to their negative electric charges. Additionally, a higher chemical potential for the $d$ and $s$ quarks increases the abundance of the negatively charged particles relative to the positively charged particles, thereby enhancing the negative values of $S_d$ and $S_s$. This behavior is observed in both isotropic and anisotropic media. 

The difference in electric charge between $u$ and $d$ quarks leads to a notable disparity in their Seebeck coefficients, with the magnitude of Seebeck coefficient for the $d$ quark being approximately twice that 
of the $u$ quark. The identical negative charge of the $d$ and $s$ quarks accounts for the close similarity 
in their Seebeck coefficients. The positive sign of $S_u$ indicates that the direction of the induced electric field points along the direction of the temperature gradient, whereas the negative sign of $S_d$ and $S_s$ implies that the directions of the induced electric field and the temperature gradient are opposite. Although both $S_d$ and $S_s$ are negative, the relative magnitudes of $S_u$, $S_d$, and $S_s$ lead to a total Seebeck coefficient for the medium that is positive, albeit small in magnitude. 

Figures \ref{Fig.3} and \ref{Fig.4} depict the effects of partonic interactions on the Seebeck coefficients for individual quark flavors and for entire QGP medium. In all cases, the inclusion of partonic interactions enhances the magnitudes of the Seebeck coefficients. The only distinction between the ideal case, corresponding to the current quark mass scenario, and the interaction case, represented by the quasiparticle mass scenario, lies in this enhancement. The increased Seebeck coefficient in the quasiparticle mass scenario signifies the role of effective thermal quark masses in governing the charge transport within the medium. In the current quark mass scenario, quarks retain their intrinsic masses, which are relatively small (e.g., $m_u \simeq 3$ MeV, $m_d \simeq 5$ MeV, $m_s \simeq 100$ MeV). In contrast, within the quasiparticle framework, each quark acquires an effective mass (eq. \eqref{E.M.(Anisotropy)}) that depends on the temperature, chemical potential and anisotropy through the thermal mass (eq. \eqref{Q.P.M.Q.(Anisotropic medium)}). This effective mass modifies the quark dispersion relation. The introduction of an effective mass alters the transport properties of quarks and, consequently, affects the Seebeck coefficient. By modifying the quark distribution function, the effective mass reshapes the phase-space occupation, enhances the contribution of specific momentum modes and suppresses the contribution of high-momentum quark states. This redistribution leads to a stronger charge separation in response to a temperature gradient, thereby resulting in an enhanced Seebeck coefficient of the QGP medium. 

\begin{figure}[]
\begin{center}
\begin{tabular}{c c}
\includegraphics[width=7.4cm]{suaniso_mix.eps}&
\hspace{0.73 cm}
\includegraphics[width=7.4cm]{sdaniso_mix.eps} \\
\ \ \ \ \ \ \ (a) & \ \ \ \  \ \ \ \ \ \ \ \ \ \ (b)
\end{tabular}
\caption{The effect of partonic interactions on the Seebeck coefficient for (a) $u$ quark and (b) $d$ quark.}\label{Fig.3}
\end{center}
\end{figure}

\begin{figure}[]
\begin{center}
\begin{tabular}{c c}
\includegraphics[width=7.4cm]{ssaniso_mix.eps}&
\hspace{0.73 cm}
\includegraphics[width=7.4cm]{saniso_mix.eps} \\
\ \ \ \ \ \ \ (a) & \ \ \ \  \ \ \ \ \ \ \ \ \ \ (b)
\end{tabular}
\caption{The effect of partonic interactions on the Seebeck coefficient for (a) $s$ quark and (b) entire QGP medium.}\label{Fig.4}
\end{center}
\end{figure}

\begin{figure}[]
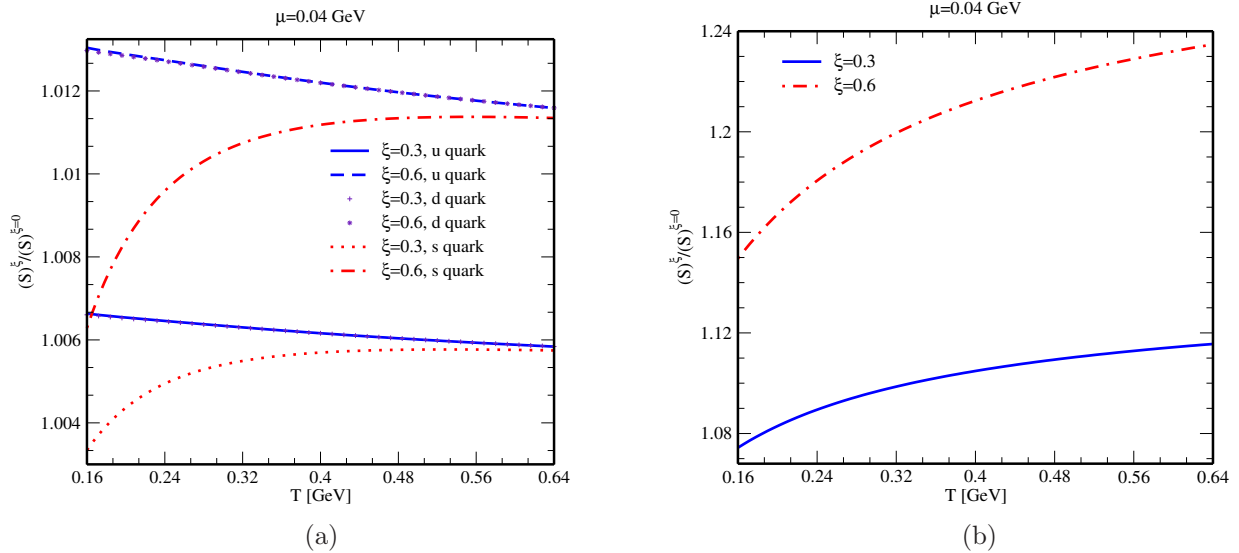

\begin{center}
\begin{tabular}{c c}
\includegraphics[width=7.4cm]{ratiocp4f.eps}&
\hspace{0.73 cm}
\includegraphics[width=7.4cm]{ratiocp4.eps} \\
\ \ \ \ \ \ \ (a) & \ \ \ \  \ \ \ \ \ \ \ \ \ \ (b)
\end{tabular}
\caption{Temperature dependence of (a) the Seebeck coefficients of different quark flavors and (b) the total Seebeck coefficient, normalized to their isotropic values, for different values of the anisotropy parameter at fixed chemical potential.}\label{Fig.5}
\end{center}
\end{figure}

Figure \ref{Fig.5}(a) shows that the deviations of Seebeck coefficients for $u$ and $d$ quarks from their corresponding isotropic counterparts decrease with the increasing temperature, in contrast to the 
increase of deviation in case of the $s$ quark. This indicates that the anisotropy has a greater impact on the thermoelectric currents of $u$ and $d$ quarks at lower temperatures, while its effect on the thermoelectric current of $s$ quark is more prominent at higher temperatures. According to figure \ref{Fig.5}(b), the deviation of the total Seebeck coefficient for the QGP medium from its isotropic counterpart increases with the growing temperature, thus corroborating the fact that the thermoelectric behavior of the medium is affected 
by the anisotropy to a greater degree at higher temperatures. 

The enhancement of the Seebeck coefficient in the presence of expansion-induced anisotropy has significant physical implications for the thermoelectric response and the transport properties of the anisotropic QGP medium. This can be comprehended from the fact that the emergence of the aforesaid anisotropy modifies both the effective quark mass and the parton distribution functions. Consequently, it shifts the phase-space occupation and increases the contribution of certain momentum modes, leading to an enhanced Seebeck coefficient. Additionally, the increase of the effective quark mass in the presence of anisotropy suppresses the contribution of high-momentum quark states. This effect results into a stronger charge separation in 
response to a temperature gradient, thus increasing the Seebeck coefficient of the QGP medium. 

The increase of the Seebeck coefficient in an anisotropic QGP medium describes a growing tendency for the medium to generate an electric field in response to a temperature gradient, suggesting a more 
efficient thermoelectric conversion in the anisotropic medium as compared to the isotropic medium. Thus the emergence of anisotropy makes the medium more efficient at converting temperature gradients into electric currents. This means that, even modest temperature gradients in the anisotropic medium can generate 
relatively stronger electric fields or charge separations than in the isotropic medium. This efficiency 
of the thermoelectric conversion in the anisotropic medium becomes larger at higher chemical potentials. 

Thermoelectric currents contribute to the electromagnetic radiation emitted from the QGP, such as soft photons and low-mass dileptons. A higher Seebeck coefficient enhances these thermoelectric currents, potentially leaving observable imprints in the electromagnetic spectra emitted during early QGP stages. The Seebeck coefficient is sensitive to both the microscopic structure and collective behavior of the medium. Its anisotropy-induced enhancement provides a potential diagnostic tool for identifying anisotropic stages in QGP evolution and for distinguishing between the isotropic and the nonequilibrium transport regimes. 

\section{Conclusions}\label{sec:conclusions}
In this work, we studied how the expansion-induced anisotropy modifies the thermoelectric properties of the QGP medium by calculating the Seebeck coefficient in this regime. We determined the Seebeck coefficients of individual quark species as well as that of the entire QGP medium by solving the relativistic Boltzmann transport equation in the relaxation time approximation within the kinetic theory approach, where the interactions among partons were incorporated through their distribution functions in the quasiparticle model at finite temperature, anisotropy and baryon asymmetry. 

We found that the magnitude of the Seebeck coefficient decreases with increasing temperature, in contrast to its growth with increasing chemical potential. Our observation further showed that the anisotropy leaves a marginal impact on the Seebeck coefficients of individual quark flavors, whereas its influence on the total Seebeck coefficient is more conspicuous. The increase in the Seebeck coefficient due to the introduction of anisotropy is mainly attributed to the enhanced parton distribution functions in the anisotropic medium. 

This increase in the Seebeck coefficient has phenomenological significance in heavy-ion collisions. In the initial stages of heavy-ion collisions, large temperature gradients are produced. An increased Seebeck coefficient leads to a stronger separation of charges due to the thermoelectric forces in such collisions. This may imply observable signals in experiments, such as charge asymmetries in particle distributions. Additionally, the enhanced Seebeck coefficient reflects changes in the quark content and dynamics as well as in the interaction strength, {\em i.e.} more interactions may enhance diffusion processes in the medium. It may also signal proximity to critical points in the QCD phase diagram or changes in the effective degrees of freedom. Furthermore, an amplified Seebeck coefficient can lead to a modified electrical conductivity and an enhanced diffusion current within the medium. Consequently, this affects the transport behavior, and potentially provides experimental signatures of the internal structure and phase properties of the QGP medium. 

\section{Acknowledgments}
One of the authors (S. R.) acknowledges financial support from ANID Fondecyt Postdoctoral Grant 3240349 
for this work. N. N. acknowledges support from ANID (Chile) FONDECYT Iniciaci\'on Grant No. 11230879.

\end{document}